\documentclass[11pt,twoside]{article}

\usepackage{newpasp}
\usepackage{graphicx}
\usepackage{lscape}

\begin{document}
\title{The Outer Structure of Galactic Disks:\\
Connections Between Bars, Disks, and Environments}   %%% Fill in title
\author{Peter Erwin}
\affil{MPI f\"ur extraterr. Physik, Giessenbachstr., 85748 Garching, Germany}    %%% Fill in author affiliations
\author{Michael Pohlen}
\affil{Astronomical Institute, University of Groningen, PO Box 800,
NL-9700 AV Groningen, The Netherlands}
\author{John E. Beckman, Leonel Guti\'{e}rrez, and Rebeca Aladro}
\affil{Instituto de Astrof\'\i sica de Canarias, E-38205 La Laguna, Tenerife,
Spain}
\begin{abstract} %%% Abstract to run on from here.
Surface-brightness profiles for early-type (S0--Sb) disks exhibit three main
classes (Type I, II, and III).  Type II profiles are more common in barred
galaxies, and most of the time appear to be related to the bar's Outer
Lindblad Resonance.  Roughly half of barred galaxies in the field have Type II
profiles, but almost none in the Virgo Cluster do.  A strong
\textit{anti}correlation is found between Type III profiles
(``antitruncations'') and bars: Type III profiles are most common when there
is no bar, and least common when there is a strong bar.
\end{abstract}

%%% MAIN BODY OF TEXT GOES HERE. CONSULT "INSTRUCTIONS FOR AUTHORS USING
%%% LATEX2E MARKUP", SECTIONS 2.3-2.6 FOR HELP WITH EQUATIONS, FIGURES,
%%% AND TABLES.

%\section{}   %%% Top level section head (remove "%" symbol)
%\subsection{}   %%% Second level section head (remove "%" symbol)
%\subsubsection{}   %%% Lowest level section head (remove "%" symbol)
%\section*{}    %%% Unnumbered top level section head (remove "%" symbol)
%\subsection*{}   %%% Unnumbered second level section head (remove "%" symbol)

\section{Introduction}

Recent imaging studies of the disks of S0 and spiral galaxies have
demonstrated that not all stellar disks have simple exponential
surface-brightness profiles (e.g., \nocite{erwin05}Erwin, Beckman, \& Pohlen
2005; \citealt{pohlen-trujillo}; \citealt{hunter06}; \nocite{erwin07}Erwin,
Pohlen, \& Beckman 2007).  Instead, disk profiles appear to fall into three
general categories: single-exponential \citet{freeman70} Type I; Freeman Type
II, with a shallow inner slope and a steeper outer slope (including so-called
``truncations''); and Type III (``antitruncations''), with a steep inner slope
and a shallow outer slope \citep{erwin05}.  See Pohlen et al.\ (this volume)
for more background and illustrations of the three types.

We report here on analysis of a deep imaging study of lenticulars and
early-type spirals (Hubble types S0--Sb), focused on tracing the outer disk
structure using azimuthally averaged surface-brightness profiles.  The 
analysis of the barred-galaxy subsample (66 galaxies) is complete 
\citep{erwin07}; analysis of the 45 \textit{unbarred} galaxies is nearly 
complete (see Aladro et al., this volume).

Pohlen et al.\ (this volume) discuss the general trend of these disk types
versus Hubble type, including the complementary late-type sample of
\citet{pohlen-trujillo}; here, we highlight what we can learn about profiles
in the earlier types and their connections to bars and galaxy environment.

\section{The Nature of Type II Profiles}

For barred galaxies, we can use the size of the bar as a measuring rod.  Type
II profiles (downward-bending broken exponentials) fall into two fairly
distinct categories: those with the break inside the bar (Type II.i) and those
with the break outside (Type II.o).  The first category is rare (6\% of the
barred galaxies), but does match profiles of some $N$-body simulations
\citep[e.g.,][]{athan02,valenzuela03}, so we suspect that this is a
side-effect of the bar formation process.

Type II.o profiles are much more common (42\% of barred galaxies), but more
difficult to explain.  One possibility is ``truncations'' due to a threshold
in star formation.  \citet{elmegreen06} recently showed that such thresholds
could produce broken-exponential Type II profiles.  However, the threshold
(gas) surface densities predicted by such models (e.g., 3--10 $M_{\sun}$
pc$^{-2}$ in \citeauthor{schaye04} 2004) are quite low.  While this is
plausibly consistent with some breaks in late-type galaxies
\citep[e.g.,][]{pohlen02}, the surface brightnesses at the break radius which
we observe correspond to stellar mass densities which an order of magnitude or
more higher: 20--2000 $M_{\sun}$ pc$^{-2}$ (median $\sim 100 \, M_{\sun}$
pc$^{-2}$).

\begin{figure}[!ht]

\includegraphics[scale=0.6]{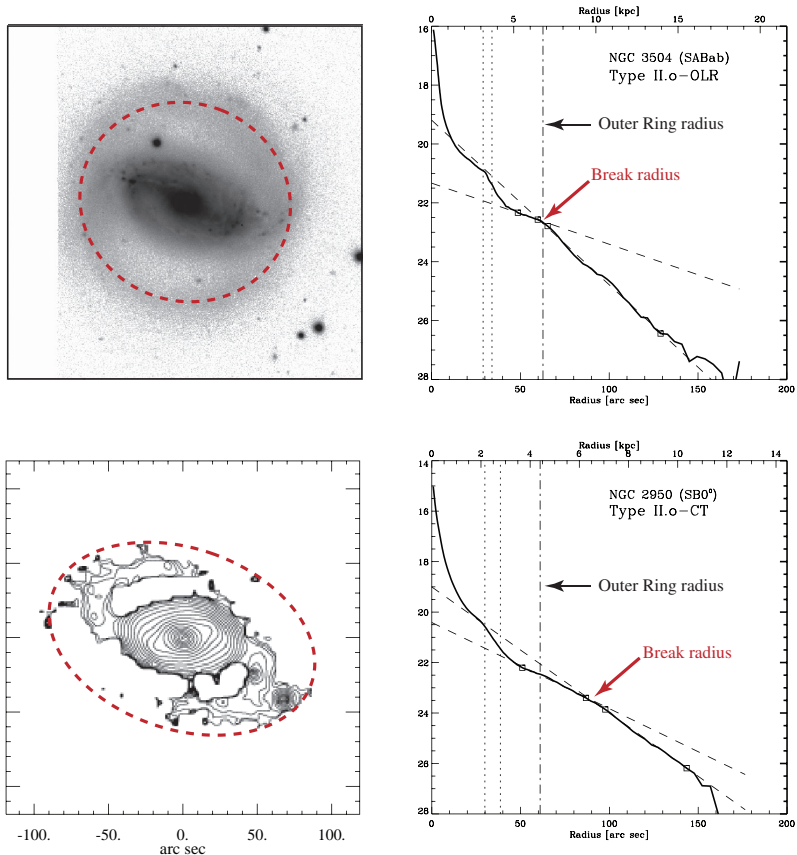}
\caption{OLR breaks (II.o-OLR) versus classical truncations (II.o-CT).  Top: a
Type II.o profile where the break coincides with an outer ring; bottom: a Type
II.o profile where the break is well \textit{outside} the outer ring.  In both
cases, the break radius is indicated by the dashed ellipse (left-hand
panels) and the arrow (right-hand panels).  For NGC~3504, we display an
SDSS $r$-band image; for NGC~2950, we have subtracted a model of the outer
disk from the SDSS image in order to bring out the (faint) outer pseudo-ring.
\label{fig:type2olr-vs-ct}}

\end{figure}

One potential clue is that we often find outer rings in Type II.o galaxies
with the ring located at or very close to the break
(Figure~\ref{fig:type2olr-vs-ct}).  Since outer rings are well understood as
an effect of a bar's Outer Lindblad Resonance \citep[OLR; see,
e.g.][]{buta96}, we suspect that these breaks are a related phenomenon.
Supporting evidence comes from recent $N$-body simulations by
\citet{debattista06}, who found that Type II.o profiles could form at the OLR
of a bar-driven spiral.  This leads us to suggest that the breaks in most Type II.o 
profiles are OLR-related, and we refer to these as ``Type II.o-OLR'' profiles.

\begin{figure}[!ht]

\includegraphics[scale=0.6]{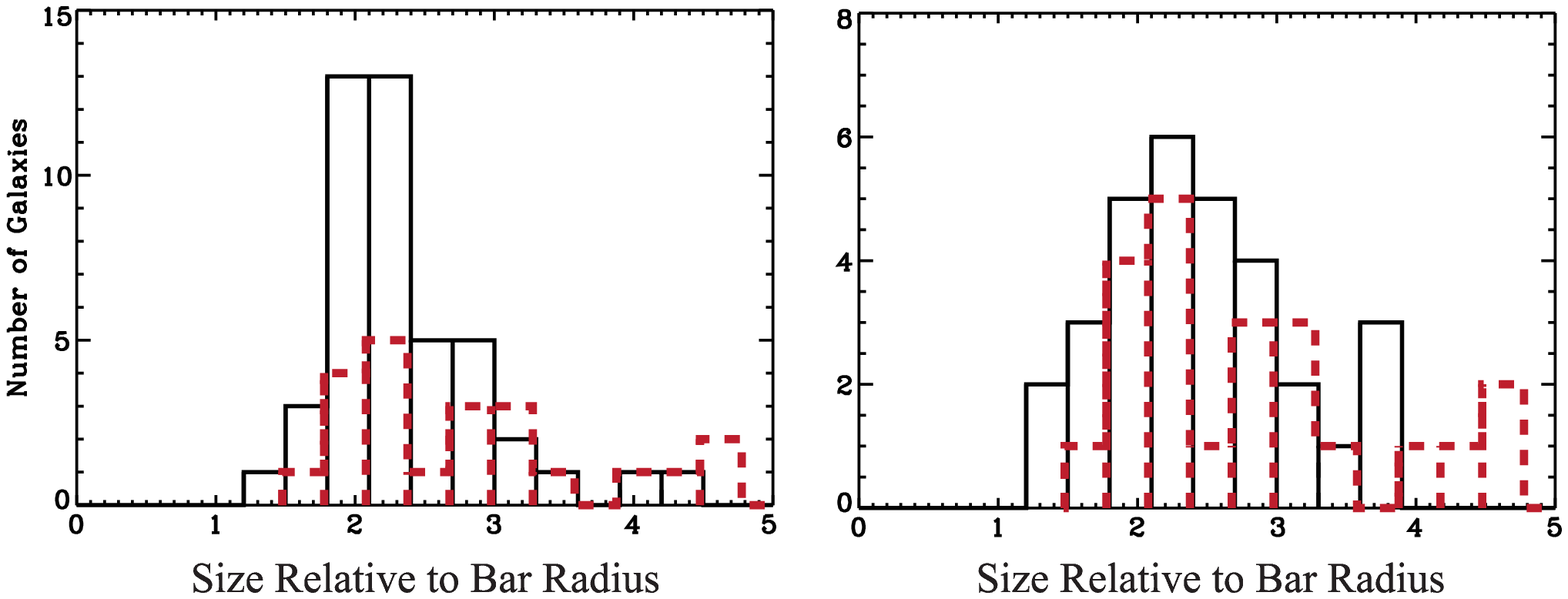}
\caption{Comparisons of outer ring sizes and Type II.o break radii.  Left:
outer-ring sizes from \citet{bc93}, using inner-ring size as a proxy for bar
radius, along with break radii from our sample (dashed histogram).  Right: As
before, except using outer-ring sizes from our sample \citep{erwin07}.  In
both cases, II.o break radii share the same size distribution as outer rings
(except for a tail at large radii), suggesting that these breaks are, like
outer rings, related the outer Lindblad resonances of bars.
\label{fig:histogram}}

\end{figure}

Figure~\ref{fig:histogram} demonstrates that the break radii in our Type II.o
profiles follow the same size distribution (break radius in terms of bar
radius) as the outer rings in our sample, as well as the outer rings in the
sample of \citet{bc93}.  A handful of galaxies have breaks at larger radii ($>
3 \times R_{\rm bar}$) --- corresponding to low stellar mass densities ---
which suggests that the star-formation threshold mechanism might be at work in
these galaxies, instead of an OLR-related mechanism.
Figure~\ref{fig:type2olr-vs-ct} shows one galaxy where this may be the case:
the break is located well \textit{outside} the bar's outer ring, and is thus
\textit{not} connected with the OLR. We label such profiles ``classical
truncations'' (Type II.o-CT).

Note that the CT profiles are rare in the early type galaxies (5\% of the
barred galaxies, 27\% of the unbarred galaxies); however,
\citet{pohlen-trujillo} found that they are more common in late type spirals
(see Pohlen et al., this volume).

% \section{Type III-d vs III-s}
% Maybe optional, depending on how much space it takes.
% Quick reference to Gil de Paz et al.?

\section{Frequencies of Disk Types versus Bars and Environments}

A striking contrast emerges when barred and unbarred galaxies are compared:
unbarred galaxies are twice as likely to have Type III profiles (60\% of
unbarred galaxies, 32\% of barred galaxies).  This carries over into the
barred galaxies themselves: weakly barred (SAB) galaxies are more likely to
have Type III profiles than strongly barred (SB) galaxies.  Similarly, the
distribution of bar strengths (deprojected maximum isophotal ellipticity) is
different for Type III and non-Type III profiles: bars in Type III galaxies
have a median (deprojected) ellipticity of $\approx 0.35$, versus $\approx
0.50$ for Types I and II.

Type II profiles (of all kinds) are clearly more common in barred galaxies
(49\% of barred galaxies versus 27\% of unbarred galaxies).  However, they
seem to be relatively unaffected by bar \textit{strength}: for example, the
Type II.o frequency is essentially identical for both SB and SAB galaxies, and
there is no difference in bar ellipticity between Type I and II.o profiles.

What \textit{does} appear to affect the presence of the Type II.o profiles is
the \textit{environment}.  Specifically, barred galaxies in the Virgo Cluster
have a much lower frequency of Type II.o profiles (10\%) than do barred
galaxies in the field (49\%).  Most of the ``missing'' II.o profiles in the
Virgo Cluster are Type I profiles instead.  Although the numbers are
relatively small, the result is significant at the 99.8\% level.  \textit{Why}
this should be so is hard to say, particularly since we don't yet know why
some field galaxies have Type I profiles and others have Type II. One
speculative possibility is that Type II.o-OLR profiles (which are the vast
majority of the field II.o profiles) require several Gyr of interaction
between the bar and gas in the outer disk.  If the gas is removed rapidly
enough, as might be the case for ram-pressure stripping of spirals which fell
into the Virgo Cluster several Gyr ago, then the OLR break formation process
might be choked off before it can proceed to completion.

\acknowledgements %%% Text of acknowledgements runs on after this command.
P.E. was supported by DFG Priority Program 1177; this work was also supported by grants No.
AYA2004-08251-CO2-01 from the Spanish Ministry of Education and Science and
P3/86 of the Instituto de Astrofisica de Canarias.

%%% THE BIBLIOGRAPHY
%%%
%%% CONSULT SECTION 3 OF "INSTRUCTIONS FOR AUTHORS" FOR HOW TO USE NATBIB.
%%% AUTHORS ARE ENCOURAGED TO USE EITHER THE "THEBIBLIOGRAPY" ENVIRONMENT
%%% BY UNCOMMENTING (DELETING THE "%" SYMBOL) THE COMMANDS BELOW, OR BY
%%% USING THE BIBTEX ENVIRONMENT. TO FIND OUT WHICH IS APPLICABLE TO YOUR
%%% CONTRIBUTION, CONSULT THE VOLUME EDITORS FOR YOUR PROCEEDINGS.
%%%

\end{document}